# Need for Objective Task-based Evaluation of Deep Learning-Based Denoising Methods: A Study in the Context of Myocardial Perfusion SPECT

Zitong Yu[1], Md Ashequr Rahman[1], Richard Laforest[2], Thomas H. Schindler[2], Robert J. Gropler[2], Richard L. Wahl[2], Barry A. Siegel[2], and Abhinav K. Jha[1, 2]

[1]Department of Biomedical Engineering, Washington University in St. Louis, St. Louis, MO, USA
[2]Mallinckrodt Institute of Radiology, Washington University in St. Louis, St. Louis, MO, USA

**Correspondence to:**
Abhinav K. Jha, PhD
Assistant Professor of Biomedical Engineering and of Radiology
Department of Biomedical Engineering
Mallinckrodt Institute of Radiology
Washington University in St. Louis
510 South Kingshighway Boulevard, St. Louis, Missouri, 63110
Telephone: +1 314-273-2655
E-mail: a.jha@wustl.edu

**Financial support:**
This work was supported in part by National Institute of Biomedical Imaging and Bioengineering of National Institute of Health (NIH) under grant number R21-EB024647, R01-EB031051, R01-EB031051-02S1, and R56-EB028287. Support is also acknowledged from the NVIDIA GPU grant. We thank Dr. Paul Segar and Duke University for providing the XCAT phantom. We also thank the Washington University Center for High Performance Computing for providing computational resources. The center is partially funded by NIH grants 1S10RR022984-01A1 and 1S10OD018091-01.

**Word count:** ~ 9000 words

**Running title**: Task-based Evaluation AI cardiac SPECT

# ABSTRACT

**Background:** Artificial intelligence-based methods have generated substantial interest in nuclear medicine. An area of significant interest has been the use of deep-learning (DL)-based approaches for denoising images acquired with lower doses, shorter acquisition times, or both. Objective evaluation of these approaches is essential for clinical application.

**Purpose**: DL-based approaches for denoising nuclear-medicine images have typically been evaluated using fidelity-based figures of merit (FoMs) such as root mean squared error (RMSE) and structural similarity index measure (SSIM). However, these images are acquired for clinical tasks and thus should be evaluated based on their performance in these tasks. Our objectives were to: (1) investigate whether evaluation with these FoMs is consistent with objective clinical-task-based evaluation; (2) provide a theoretical analysis for determining the impact of denoising on signal-detection tasks; and (3) demonstrate the utility of virtual clinical trials (VCTs) to evaluate DL-based methods.

**Methods:** A VCT to evaluate a DL-based method for denoising myocardial perfusion SPECT (MPS) images was conducted. To conduct this evaluation study, we followed the recently published best practices for evaluation of AI algorithms for nuclear medicine (the RELAINCE guidelines). An anthropomorphic patient population modeling clinically relevant variability was simulated. Projection data for this patient population at normal and low-dose count levels (20%, 15%, 10%, 5%) were generated using well-validated Monte Carlo-based simulations. The images were reconstructed using a 3-D ordered-subsets expectation maximization-based approach. Next, the low-dose images were denoised using a commonly used convolutional neural network-based approach. The impact of DL-based denoising was evaluated using both fidelity-based FoMs and area under the receiver operating characteristics curve (AUC), which quantified performance on the clinical task of detecting perfusion defects in MPS images as obtained using a model observer with anthropomorphic channels. We then provide a mathematical treatment to probe the impact of post-processing operations on signal-detection tasks and use this treatment to analyze the findings of this study.

**Results**: Based on fidelity-based FoMs, denoising using the considered DL-based method led to significantly superior performance. However, based on ROC analysis, denoising did not improve, and in fact, often degraded detection-task performance. This discordance between fidelity-based FoMs and task-based evaluation was observed at all the low-dose levels and for different cardiac-defect types. Our theoretical analysis revealed that the major reason for this degraded performance was that the denoising method reduced the difference in the means of the reconstructed images and the channel operator-extracted feature vectors between the defect-absent and defect-present cases.

**Conclusions**: The results show the discrepancy between the evaluation of DL-based methods with fidelity-based metrics vs. the evaluation on clinical tasks. This motivates the need for objective task-based evaluation of DL-based denoising approaches. Further, this study shows how VCTs provide a mechanism to conduct such evaluations computationally, in a time and resource-efficient setting, and avoid risks such as radiation dose to the patient. Finally, our theoretical treatment reveals insights into the reasons for the limited performance of the denoising approach and may be used to probe the effect of other post-processing operations on signal-detection tasks.

# INTRODUCTION

Artificial intelligence-based methods, such as those based on deep learning (DL), have generated substantial interest in nuclear medicine. An area of significant interest has been using DL-based approaches for predicting normal-dose images from images acquired at lower doses, an operation referred to as denoising. Objective evaluation of these approaches is essential for clinical translation. Medical images are acquired for specific clinical tasks. Thus, evaluation of these approaches should be based on the clinical task for which imaging was performed. Multiple methods have been developed to perform this task-based assessment of image quality[1-3], and the efficacy of this evaluation procedure has been demonstrated in multiple studies.[4-10]

Currently, DL-based denoising methods for medical imaging are typically evaluated using figures of merit (FoMs) such as root mean squared error (RMSE) and structural similarity index measure (SSIM).[11-13] These FoMs measure fidelity between the images obtained using DL-based denoising approaches as compared to some reference images. However, it is unclear if evaluation with these FoMs necessarily correlates with performance on clinical tasks that are required from these images.[14] Several recent studies have observed the limitations of these FoMs.[15-20] In a study evaluating a DL-based denoising approach for CT images, KC et al. observed discrepancies between interpretation yielded by fidelity-based FoMs vs. that obtained using bench testing metrics such as noise power spectrum.[15] In another study, Li et al. observed similar limitations when evaluating a DL-based denoising method on binary signal-detection tasks with stylized numerical studies, more specifically, a synthetic two-dimensional lumpy background-based model and an idealized planar imaging system.[16] Michael et al. observed no significant correlation between peak signal-to-noise ratio and SSIM values and classification performance for a tumor classification task in chest radiographs.[19] A study by our group observed these limitations with a 2D single photon emission computed tomography (SPECT) system with lumpy background-based tracer distribution models.[21] While these studies show the limitations of these FoMs, it is unclear if the results from these studies are indicators of performance in clinical settings in nuclear medicine. For example, while lumpy backgrounds have useful mathematical properties, they may have limitations in modeling human anatomy and physiology for clinical nuclear-medicine studies. Results for planar or 2D tomography systems may not generalize to realistic 3D SPECT systems. Similarly, following clinical reconstruction and post-processing protocols is known to improve the performance of observers in nuclear medicine on detection tasks.[22,23] Given the promise of DL-based methods in nuclear-medicine applications and the typical evaluation of these DL-based methods with fidelity-based FoMs, there is a strong need to investigate the concordance between fidelity-based FoMs vs. task-based measures of performance in clinically realistic settings for nuclear-medicine applications.

To evaluate the concordance between fidelity-based FoMs and task-based FoMs for AI-based nuclear-medicine imaging methods, we considered the evaluation of a DL-based method for denoising myocardial perfusion SPECT (MPS) images acquired at low dose using both these FoMs. There has been significant interest in using DL-based methods for denoising low-dose SPECT images.[24-27] We investigated whether the evaluation of one such commonly used DL-based denoising method using fidelity-based FoMs was concordant with the evaluation of these methods on the task of detecting myocardial perfusion defects, a major clinical task for which these images are acquired. For the evaluation to be clinically realistic, we conducted this study as a virtual clinical trial (VCT) (also referred to as in silico imaging trial or virtual imaging trial).[28,29] VCTs offer a mechanism to evaluate new medical imaging technologies virtually.[30] In a VCT, the human population is replaced with a digital phantom population, the imaging system by a simulated scanner, and the clinical interpretation by a model observer. Further, because the ground truth is known in a VCT, evaluation can be performed on clinically relevant tasks. Thus, our VCT enabled the evaluation of the DL-based denoising method in a clinically realistic situation and on a clinically relevant task.

We also provide a detailed theoretical analysis to quantify the impact of post-processing operations, such as denoising, on signal-detection tasks with observers that use the first and second-order statistics of the images. This includes the channelized Hotelling observer (CHO)[14,31-33], which we use as the model observer in our VCT. Thus, we use our theoretical analysis to provide insights on the results obtained in the VCT on evaluating the DL-based denoising method on the task of detecting myocardial perfusion defects.

Finally, another major objective of this research is to demonstrate the utility of VCTs for task-based evaluation of AI algorithms for nuclear medicine. Clinical translation of AI algorithms for nuclear medicine requires rigorous evaluation on clinical tasks. However, conducting clinical evaluation studies is expensive, time-consuming, exposes patients to radiation dose, and suffers from multiple logistical challenges such as requiring patient recruitment and consent for prospective trials and availability of trained human observers. VCTs offer a mechanism to address these challenges by providing the ability to simulate population variability, model imaging system physics, and because the ground-truth is known, the ability to perform evaluation on detection and quantification tasks. Thus, a VCT-based evaluation can be used to identify promising methods for subsequent clinical evaluations. Through this study, we demonstrate the applicability, utility, and feasibility of conducting evaluation on clinical tasks in clinically realistic settings using the VCT framework.

## METHODS

In this study, we used a VCT to evaluate the performance of a DL-based denoising method for low-dose MPS on the task of cardiac defect detection. More specifically, we conducted this study with an anthropomorphic phantom-based population that simulated patient-population variability, accurately modeled the SPECT system using well-validated Monte Carlo-based software, and used validated anthropomorphic model observers that have been shown to emulate the performance of human observers for the task of cardiac defect detection from MPS images. To conduct the evaluation study, we followed the recently published best practices for evaluation of AI algorithms for nuclear medicine (the RELAINCE guidelines)[34]. A schematic of the workflow of our VCT is shown in Fig. 1. We now describe the individual components of the study.

**Generating the digital-phantom population**

Digital activity and attenuation maps of the thoracic region were generated using the anthropomorphic 3-D extended cardiac and torso (XCAT) phantom software. This software provides a mechanism to generate phantoms with highly detailed anatomies,[35] simulate physiological and anatomical variability as observed in patient populations,[36] and produce realistic imaging data when combined with an accurate model of the imaging process.[35] Further, the procedure we followed to generate our population was a procedure outlined specifically to generate phantom populations for MPS research,[37] and has since been used in several realistic MPS simulation studies.[9,38-40] Following this specific procedure strengthened the clinical realism of the generated phantom population, as we describe below.

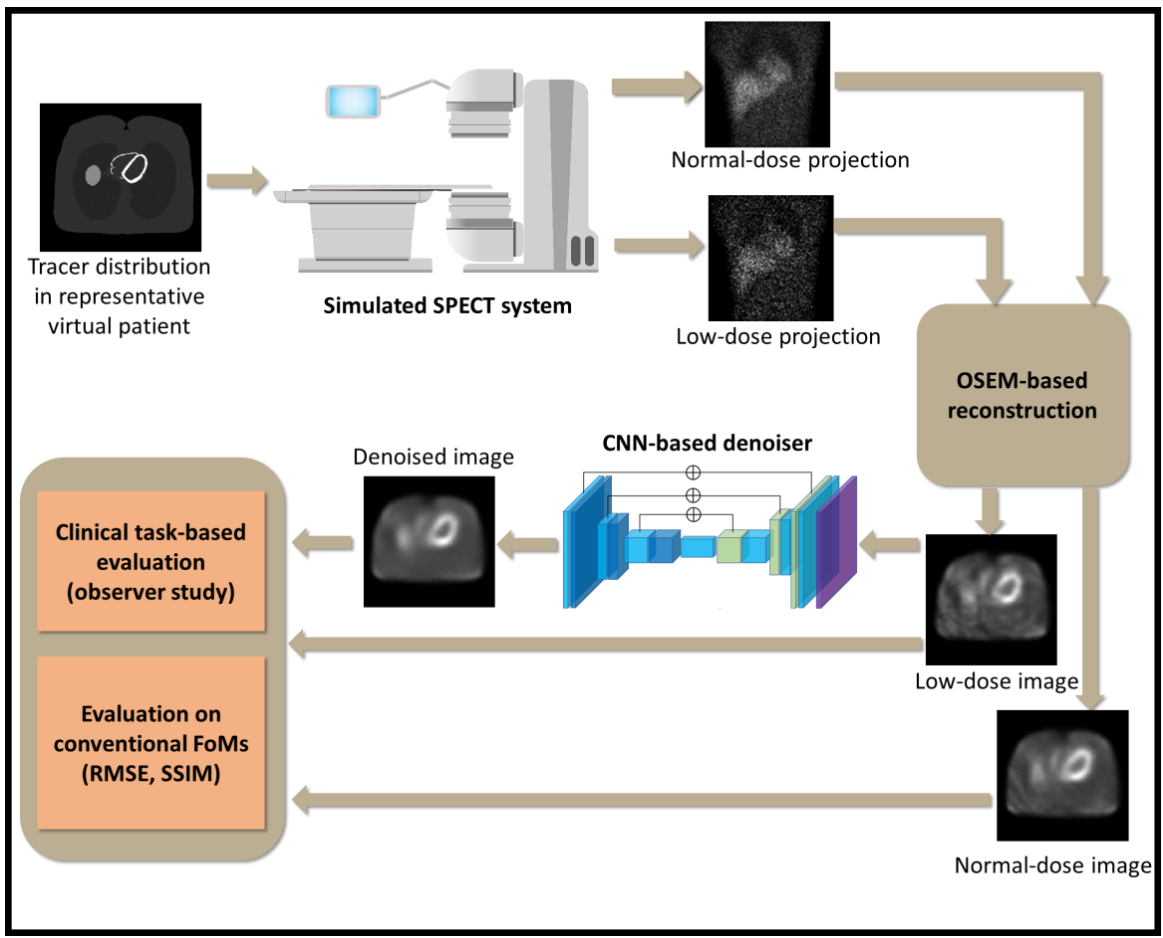

*Figure 1: The workflow of our virtual clinical trial.*

We generated a virtual population of N = 4,800 patients, including both diseased and healthy patients. The population consisted of both male and female patients in equal numbers and had variations in both anatomy and physiology. The anatomical parameters of the patient for the male and female populations are presented in Table I. Different body sizes, left ventricle (LV) length and radius for each of the 4,800 patients were randomly sampled from these truncated normal distributions. The skin-layer thickness of the 3D XCAT phantoms were scaled by the factors 0.5 to 1.5 of the average skin layer thickness for the given body size to model the thickness of the subcutaneous adipose tissue. To model realistic body shapes and heart shapes for male and female population, we considered the ratio of the body lateral to anteroposterior dimensions fixed to 1.36 and 1.47, respectively, and the LV length-to-radius ratio fixed to 3.2 and 3.17, respectively, as in the Emory Cardiac Database.[41]

TABLE I: Anatomical parameters of the patient population

|  |  | Body LAT[a,e] | Body AP[b] | LV[c] Length | LV Radius | Height |
|---|---|---|---|---|---|---|
| Male | Mean | 34.84 | 25.70 | 8.31 | 2.67 | 175.68 |
|  | SD[4] | 2.15 | 2.44 | 0.93 | 0.47 | 6.80 |
|  | Min | 29.40 | 20.00 | 6.60 | 1.90 | 154.94 |
|  | Max | 38.40 | 31.40 | 11.60 | 4.00 | 187.96 |
| Female | Mean | 34.37 | 23.50 | 7.39 | 2.32 | 163.45 |
|  | SD | 3.25 | 2.08 | 0.92 | 0.33 | 7.34 |
|  | Min | 26.70 | 19.60 | 5.70 | 1.60 | 149.86 |
|  | Max | 40.90 | 28.80 | 10.50 | 3.50 | 177.80 |

[a]LAT: Lateral dimension

[b]AP: Antero-posterior dimension

[c]LV: Left ventricle

[d]SD: Standard deviation

[e]All units in cm

To simulate the variability in physiology, we varied the activity uptake in the heart, lung, liver, and the rest of the thoracic region across the entire patient population. The ratios of activity in these regions were sampled from truncated-normal distributions, the parameters of which were derived from a set of 34 clinical Tc-99m-sestamibi MPS studies.[5] The parameters that describe the distribution of tracer uptake ratios are listed in Table II.

TABLE II: Parameters describing the distribution of the tracer uptake ratios in different organs

|  | Liver/Heart | Lung/Heart | Background/Heart |
|---|---|---|---|
| Mean | 0.44 | 0.14 | 0.11 |
| SD[a] | 0.19 | 0.04 | 0.05 |
| Min | 0.16 | 0.05 | 0.02 |
| Max | 1.3 | 0.25 | 0.29 |

[a]SD: Standard deviation

Next, we simulated four different clinically relevant myocardial perfusion defects. The defects were characterized by low uptake compared to the rest of the heart. Each defect had a different extent, cold contrast (referred to as contrast), or location. The defect properties were as in previous studies.[33,42] The defect extents were characterized by the width of the defect in the circumferential dimension. A width of 90° meant that the width is 45° on either side of the defect center. Defect contrast was characterized by the ratio of the uptake in the defect to healthy myocardial tissue. In the rest of the manuscript, we refer to the four types of defects as type 1, type 2, type 3, and type 4, with parameters as stated in Table III.

We split our patient population of N = 4,800 into four equal sub-populations. In each sub-population, one of the four defects was introduced in half of the patients. Thus, by the end of this process, we had four patient sub-populations, each of which had a different defect type. The phantoms were digitized into 512 × 512 × 114 3-D phantoms, with a voxel size of 0.11 cm × 0.11 cm × 0.44 cm. The workflow to generate the digital-phantom population is shown in Fig. 2.

TABLE III: Defect Parameters

| Defect Type | Contrast (%) | Extent (degrees) | Location |
|---|---|---|---|
| 1 | 50 | 120 | Anterior of LV wall |
| 2 | 25 | 90 | Anterior of LV wall |
| 3 | 50 | 120 | Inferior of LV wall |
| 4 | 25 | 90 | Inferior of LV wall |

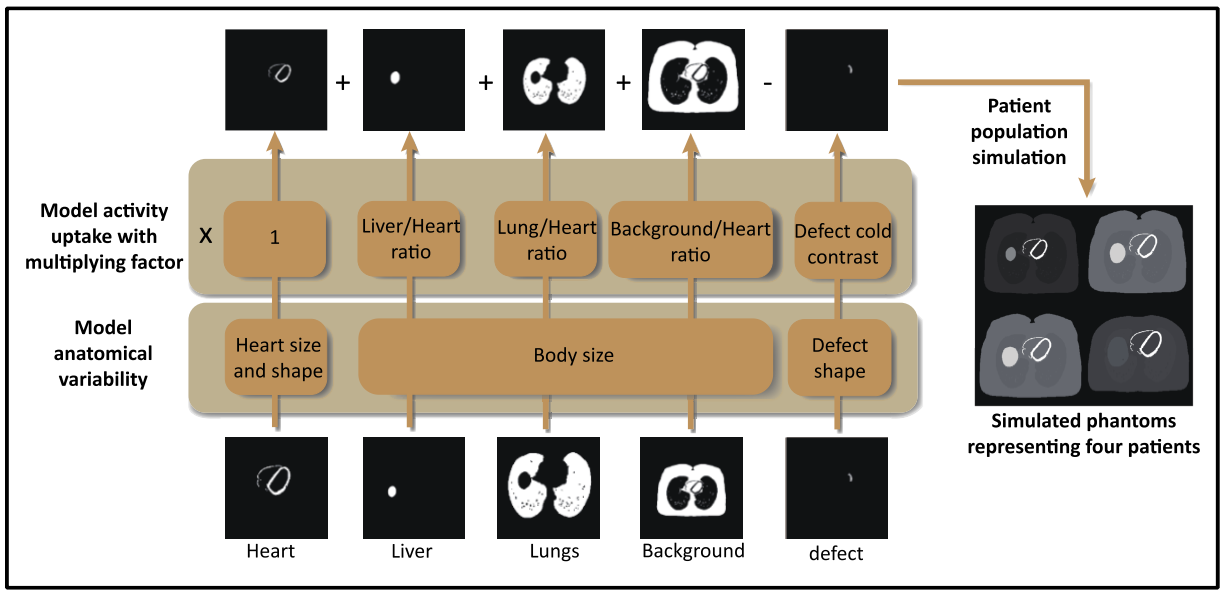

Figure 2: The workflow to generate the digital-phantom population

**Modeling the image-formation process**

To accurately generate the projection data from the 3-D phantoms, we used SIMIND,[43,44] a well-validated Monte Carlo-based software. We simulated a clinical 3-D parallel-hole system with a configuration similar to the GE Optima NM/CT 670 SPECT/CT scanner and a low-energy high-resolution (LEHR) parallel-hole collimator. The emission source was assumed to be $^{99m}$Tc, the most common tracer for MPS. We acquired projections within a 20% photopeak window centered at 140 keV at 60 angular positions spaced uniformly over 180° from left posterior oblique to right anterior oblique, modeling a constant distance orbit. The scintillation detector had an intrinsic spatial resolution of 4 mm and an energy resolution of 9.8% at 140 keV. The SPECT system had a resolution of 7.4 mm at 10 cm depth. The simulation modeled all relevant image-degrading processes in SPECT, including the scatter and resultant attenuation of the photons, the depth-dependent collimator response, and the finite energy and position resolution of the detector.

The projection data were collapsed to 0.44 cm projection bins in a 128×114 projection image matrix. We simulated the detection of a large number of photons, generating an approximately noiseless image. We then scaled these counts to clinical normal-dose level (an average of 12 million projection counts from the thoracic region) as well as 20%, 15%, 10%, and 5% of the normal-dose level. Poisson noise was added to the projection at both normal and low-dose levels. We chose these four low-dose levels so that the performance could be significantly different from that obtained at the normal dose level, thus enabling us to study whether the use of a DL-based denoising approach did indeed improve performance compared to the low-dose images. He et al. have shown that when the dose level is low, quantum noise dominates over anatomic variability in the evaluation of MPS imaging methods on defect-detection task.[45] Thus, choosing these low-dose levels yielded a regime where the quantum noise impacted task performance and where we could investigate if the DL-based denoising helped to reduce the impact of this noise on task performance.

At the end of this process, for each defect type, we generated 600 pairs of defect-present and defect-absent projections at each dose level.

**Image Reconstruction and post-processing**

We reconstructed the projection data using a clinically used 3-D ordered-subsets expectation maximization (OSEM) based technique.[46] The technique compensated for the major image-degrading processes in SPECT, including the attenuation and scatter of photons, and the collimator-detector and geometric response. As per clinical settings, we used four iterations with four subsets in the OSEM algorithm. A total of 24,000 reconstructed 3-D images ((600 defect-absent cases +600 defect-present cases) × (4 low dose levels +1 normal dose level) × 4 defect types) were generated. The images were of size 128×128×114, with a voxel size of 0.44 cm. Fig. 4 shows sample transaxial slices from reconstructed images obtained at normal-dose level. After reconstruction, the images were post processed using a low-pass filter. As in clinical settings, a 3-D Butterworth filter of order five and cutoff frequencies 0.4 cm$^{-1}$ was used. Negative values were set to zero.

Next, a DL-based approach was used to predict the normal-dose images from the low-dose images. Note that our objective in this study was not to propose a new DL-based denoising approach, but instead investigate the concordance in evaluating DL-based approaches with fidelity-based FoMs vs. on specific tasks. Given this objective, we used a very commonly used DL-based denoising approach for nuclear-medicine imaging, namely that based on convolutional neural networks (CNNs).[24,47-49] The CNN used in this study was based on a combination of a chain of convolutional layers and symmetric deconvolutional layers. The convolutional layers extracted the latent representation from the input image. Then the deconvolutional layers reconstructed the denoised image from the latent representation. After each convolutional layer, a leaky rectified linear unit (ReLU) activation function was applied. In this network, skip connections with element-wise addition were applied to feed the features learned in the convolutional part of the network into the deconvolutional part, allowing the network to learn more complex features.[50,51] Finally, the loss function used in the CNN was the mean square error (MSE), a commonly used loss function used in several previous CNN-based denoising studies.[24,49,52,53]

A total of sixteen CNNs (four defect types × four low dose levels) were trained separately in this study. Each CNN was trained on 200 pairs of normal-dose images and corresponding low-dose images for each defect type and at each dose level. The CNNs were optimized for each low-dose level separately via the Adam optimization algorithm[54] with a learning rate of 0.001 and batch size of 32. We used a grid-search strategy to optimize the architecture of the CNN, including the number of convolutional and deconvolutional blocks and the number of filters, for each defect type at each dose level. We conducted five-fold cross-validation during the grid search strategy. For each CNN, we selected the configuration that yielded the best performance in terms of MSE between the predicted and normal-dose images. The detailed architecture of the network is summarized in Table S I in the supplementary material. We implemented our network using TensorFlow 1.10.0 with Keras 2.2.4 on state-of-the-art NVIDIA V100 with 32 GB of memory.

**Evaluation of the DL-based Approach**

The impact of the DL-based denoising was evaluated using both fidelity-based FoMs, and the task of detecting cardiac defects. Each of the sixteen CNNs was evaluated using 400 pairs of test images for each defect type and at each dose level. The test images were different from the images used for training. Below we describe the evaluation studies.

*Evaluation with fidelity-based FoMs*: As mentioned above, conventionally, DL-based denoising methods are evaluated using fidelity-based FoMs, of which those most used include RMSE, SSIM, and peak signal-to-noise ratio (PSNR).[55-57] Of these, PSNR is a monotonic transformation of RMSE. Thus, we report results with the RMSE and SSIM. Computing these FoMs requires a reference image. As in previous studies,[24,47,58] the reference was the normal-dose reconstructed image.

*Task-based evaluation*: The major clinical task from the MPS images is detecting cardiac defects. We conducted an observer study to evaluate the impact of DL-based denoising on this task. This observer study was conducted for images acquired at normal dose and on the low-dose images prior to and after DL-based denoising. Preferably, such a study should be conducted by a human observer. However, human observer studies are tedious, time-consuming, and expensive, especially as we were conducting these studies with a large number of images. Thus, we instead conducted this study with an anthropomorphic model observer that has been validated to emulate human-observer performance for the considered task.[32,33]

In our study, in each of the four sub-populations, there was variation in anatomy and physiological uptake of the radiotracer, but the defect type was fixed and known. In this scenario, for each sub-population, while the defect type was known exactly, background variability was known statistically. For this task, it has been demonstrated in several studies that a CHO with rotationally symmetric frequency channels[4] emulates human-observer performance.[32,33] Wollenweber et al. have validated this emulation at one-tenth of the clinical-count levels for an MPS study.[32] Similarly, Sankaran et al. have validated this emulation at one-eighth to one-tenth of the clinical-count levels, again for an MPS application.[33] Thus, we used this CHO in our study. We used five 2-D rotationally symmetric frequency channels whose start frequency and width of the first channel were both 1/64 cycles per pixel. Subsequent channels were adjacent to the previous one and had double the channel width of the previous one.

To apply the CHO, we followed the same procedure that was used to validate this model observer in human studies. More specifically, we chose the 2-D slice that contains the longitudinal midpoint of defects. Then, we extracted regions of size 32 × 32 pixels from the image under consideration such that the centroid of the defects was at the center of the images. The gray level values in these extracted regions were scaled to lie in the range [0,255]. Next, we applied the anthropomorphic channels on the extracted regions, yielding the feature vectors. The test statistic for each of the 400 pairs of test images represented by feature vectors in the test set was calculated using a leave-one-out strategy. The test statistic was compared to a threshold to classify the image into the defect-present or defect-absent class. By varying the threshold, a receiver operating characteristics (ROC) curve was generated,[14,59] from which the area under the ROC curve (AUC) was obtained. We used LABROC4 program to estimate the ROC curves.[60] The LABROC4 program uses a binormal model to estimate the ROC curve and the AUC. The AUC quantified the performance on the task of defect detection. Further, 95% confidence intervals of the AUC were computed using a bootstrap-based strategy. We compared the AUC values obtained with the low-dose images prior to and after applying the DL-based denoising approach.

**Theoretical investigation of the Impact of Denoising**

To interpret the findings obtained from our VCT-based study, we conducted a theoretical analysis to investigate the impact of post-processing operations, such as denoising, on signal-detection tasks when performed by observers that use the first and second-order statistics of images. As per the data-processing inequality, any post-processing of images will not increase the task-based information extracted by the ideal observer for that task.[61] In other words, the ideal-observer performance will not be increased due to any post-processing procedures. However, for sub-optimal observers, such as human observers, it is possible that the DL-based denoising approach may improve observer performance. Further, as mentioned above, studies have shown that for the task that we are considering in our experiments, the performance of the human observer can be approximated by a CHO. This observer uses the first and second-order statistics of the feature vectors. Thus, we developed a theoretical eigenanalysis-based treatment that quantifies the impact of post-processing operations on first and second-order statistics on the feature vectors. In this sub-section, we first provide the background for this treatment, follow that with the eigenanalysis-based treatment, and then use this theoretical treatment to investigate the impact of the DL-based denoising approach on the signal-detection task.

*Background for the eigenanalysis-based treatment:* In our model-observer study, while the defect was known, the background varied over a large patient population. In this setting, the test statistics are normally distributed[14] and a monotonic relationship exists between the AUC and signal-to-noise ratio (SNR) of the model observer. This relationship is given by

$$AUC = \frac{1}{2} + \frac{1}{2}\text{erf}(SNR), \tag{1}$$

where $\text{erf}(z)$ is the error function.

The model observer in this manuscript, the CHO, operates on feature vectors. These feature vectors, denoted by $v$, are obtained by applying a channel operator on the reconstructed images. Assume that we have $L$ channels that are applied to the image. Then $v$ is a $N$-dimensional vector. Denote the reconstructed image by $\hat{f}$, and the channel operator by $U$. Then:

$$v = U\hat{f}, \tag{2}$$

where $v$ lies in Euclidean space $\mathbb{E}^L$. Denote the mean difference between feature vectors of signal-present and signal-absent patients by $\Delta\bar{v}$, and denote the covariance matrix of the feature vectors by $K_v$. The expression of the SNR for the CHO is given by[14]

$$SNR^2 = \Delta\bar{v}^T K_v^{-1} \Delta\bar{v}. \tag{3}$$

From Eq. 3, we notice that a change in $\Delta\bar{v}$ impacts the observer SNR. Thus, we plotted $\Delta\bar{v}$ obtained with the normal-dose images and the images prior to and after denoising. To gain more insights at the image level, we recognized that if $\Delta\bar{\hat{f}}$ were to denote the mean difference in the windowed reconstructed images between the defect-present and defect-absent cases, then due to the linearity of the channel operator, from Eq. 2, we can derive that

$$\Delta\bar{v} = U^T \Delta\bar{\hat{f}}. \tag{4}$$

Thus, we also plotted the $\Delta\bar{\hat{f}}$ between the reconstructed images obtained at normal dose and the images at low dose before and after denoising.

While studying the impact of the post-processing operation on the $\Delta\bar{v}$ term is helpful, that does not provide a complete picture since the observer SNR also depends on $K_v$. To obtain this complete picture, we conducted an eigenanalysis-based treatment as described below.

*Eigenanalysis-based treatment:* We note that $K_v$ is a Hermitian matrix of size $L \times L$. Thus, we can perform an eigenanalysis of this matrix, which will yield a set of eigenvectors that can be used to represent any vector in the space $\mathbb{E}^L$. Denote the m[th] eigenvalue of the covariance matrix by $\lambda_m$, and denote the m[th] eigenvector by $u_m$. Then the covariance matrix can be decomposed as below.

$$K_v = \sum_{m=1}^{L} \lambda_m u_m u_m^T. \tag{5}$$

Since the covariance matrix $K_v$ is Hermitian, the inverse of the covariance matrix can be expressed by the same eigenvalues and eigenvectors, as shown below.

$$K_v^{-1} = \sum_{m=1}^{L} \frac{u_m u_m^T}{\lambda_m}. \tag{6}$$

Next, note that $\Delta \bar{v}$ also lies in $\mathbb{E}^L$. Thus, we next represent $\Delta \bar{v}$ by the eigenvectors of $K_v$ as

$$\Delta \bar{v} = \sum_{m=1}^{L} \alpha_m \boldsymbol{u}_m, \tag{7}$$

where the coefficient $\alpha_m$ is given by the scalar product,

$$\alpha_m = \boldsymbol{u}_m^T \Delta \bar{v}. \tag{8}$$

Inserting the expressions in Eqs. 6 and 7 in Eq. 3, the observer SNR can be written as

$$SNR^2 = \sum_{i=1}^{L} \alpha_i \boldsymbol{u}_i^T \sum_{m=1}^{L} \frac{\boldsymbol{u}_m \boldsymbol{u}_m^T}{\lambda_m} \sum_{j=1}^{L} \alpha_j \boldsymbol{u}_j. \tag{9}$$

We can use the orthogonality of the eigen vectors to further simplify this equation as follows:

$$SNR^2 = \sum_{i=1}^{L} \sum_{m=1}^{L} \sum_{j=1}^{L} \frac{\alpha_i \alpha_j}{\lambda_m} \boldsymbol{u}_i^T \boldsymbol{u}_m \boldsymbol{u}_m^T \boldsymbol{u}_j \tag{10}$$

$$= \sum_{i=1}^{L} \sum_{m=1}^{L} \sum_{j=1}^{L} \frac{\alpha_i \alpha_j}{\lambda_m} \delta_{im} \delta_{mj}$$

$$= \sum_{m=1}^{L} \frac{\alpha_m^2}{\lambda_m},$$

where $\delta_{ij}$ is the Kronecker delta function that has a value of one when $i = j$ and zero otherwise.

From Eq. 10, we notice that the observer SNR is impacted by both $\alpha_m$ and $\lambda_m$. More specifically, the observer SNR depends directly on $\alpha_m$ and inversely on the square root of $\lambda_m$. Thus, based on this theoretical treatment, we investigated the impact of the DL-based denoising approach on $\alpha_m$ and the square root of $\lambda_m$.

*Investigation of the impact of the DL-based denoising approach on defect-detection performance:* As mentioned earlier, for each defect type, as per Eq. 3 and 4, we computed and compared the $\Delta \bar{v}$ and $\Delta \bar{\bar{f}}$ with the normal-dose images and the images prior to and after the DL-based denoising approach. Next, based on the eigenanalysis treatment, we conducted eigenvalue decomposition to the covariance matrix of feature vectors obtained with the normal-dose images and the images prior to and after denoising for each defect type (Eq. 5) and plotted the eigenvalue spectra. Also, as per Eq. 8, we computed and plotted the coefficients $\alpha_m$ for each defect type again with the normal-dose images and with the images prior to and after denoising. Finally, as per Eq. 10, we plotted the observer SNR for each of the defect types to show the combined effect of $\alpha_m$ and eigen values $\lambda_m$ on observer performance.

# RESULTS

Fig. 3 shows the performance of the DL-based approach using the fidelity-based FoMs and the model observer study. The figure shows the RMSE, SSIM, AUC values at all dose levels, and the ROC plots at 10% dose level. Similar ROC plots are observed at other dose levels, as presented in the supplementary material (Fig. S III). The specific values and corresponding confidence intervals of SSIM, RMSE, and AUC at all dose levels are listed in Table S II and Table S III in supplementary material.

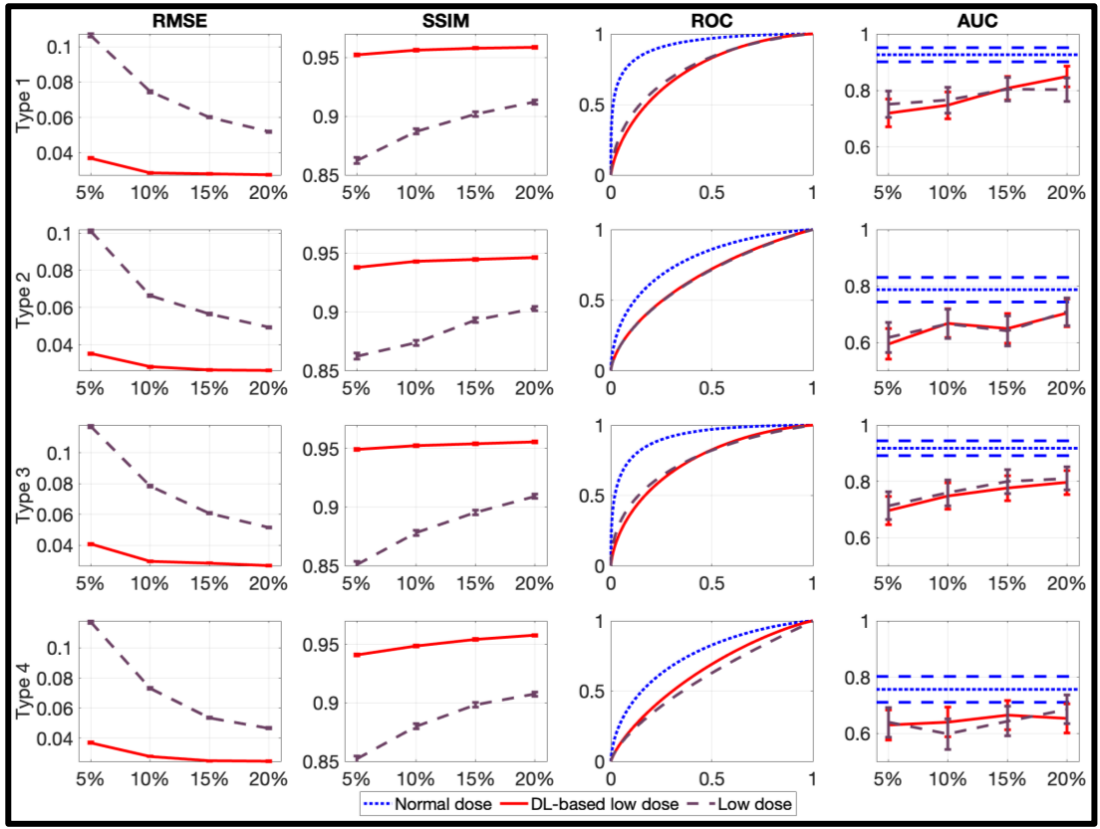

*Figure 3: The performance of the DL-based denoising approach evaluated using the fidelity-based FoMs and the model-observer study. Each row corresponds to a different defect type. The first two columns show the performance evaluated using RMSE and SSIM, respectively. The third column shows the ROC plots obtained for the normal-dose images and the low-dose image prior to and after conducting the denoising operation at the 10 % dose level. The x and y axis of this plot denote the false positive fraction (FPF) and the true positive fraction (TPF), respectively. Finally, the fourth column shows the AUC values at all the considered dose levels.*

We observe that evaluation with RMSE and SSIM FoMs leads to the inference that the DL-based denoising yields significantly superior performance compared to without denoising. This is consistently inferred for all defect types and at all dose levels. In contrast, we observe in the results from the evaluation on the defect-detection task that the ROC curves obtained using images prior to and after applying the DL-based approach almost overlapped. There was no statistically significant difference in the AUC values after applying denoising. This was true at all dose levels and for all defect types. Thus, these results lead to the inference that performing the DL-based denoising operation did not improve performance on the defect-detection task. Thus, the inferences obtained from the fidelity FoM-based and task-based interpretations were discordant.

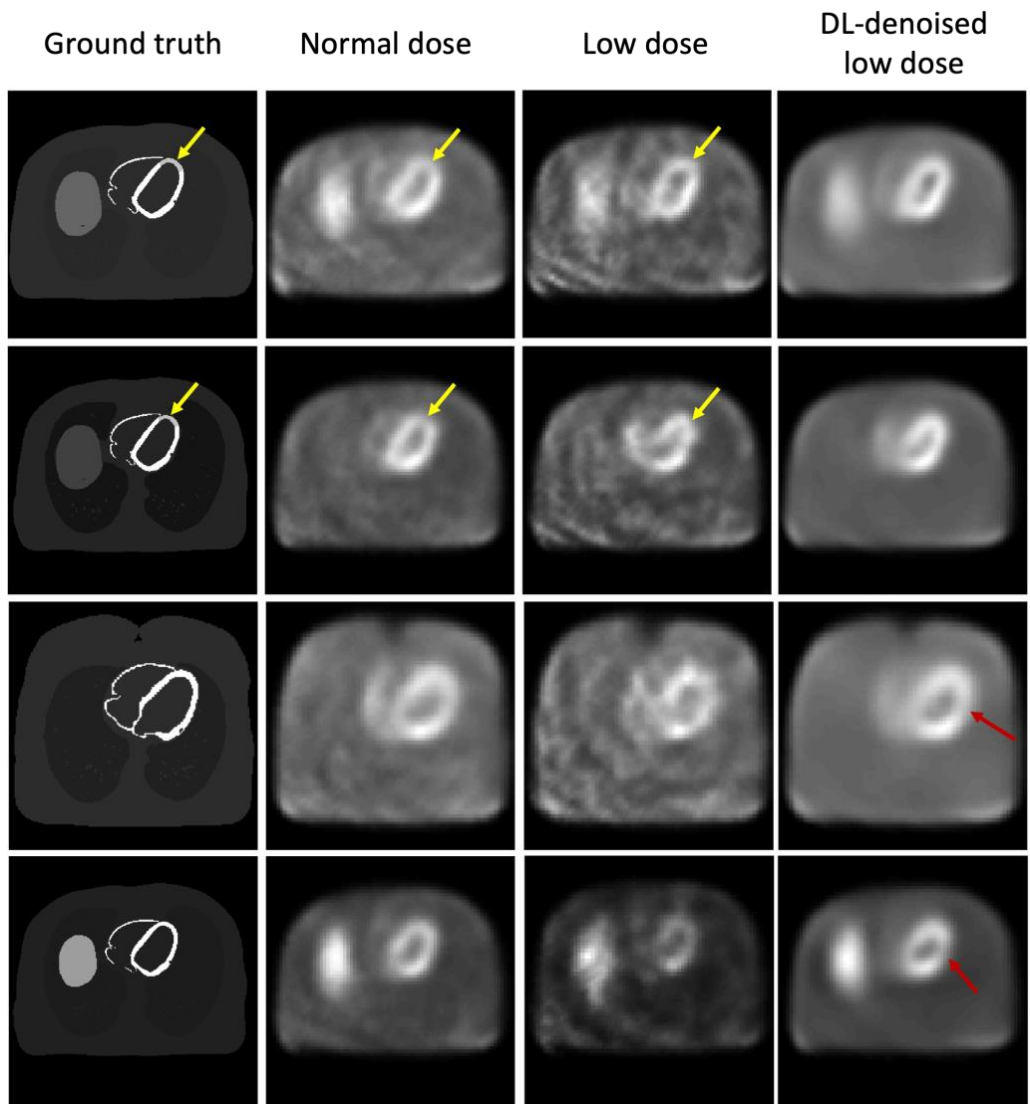

*Figure 4: Representative reconstructed images: the upper two rows show examples where the DL-based denoising approach removed the defect (defects marked by yellow arrows). Note that the defect is characterized by a lower uptake than the rest of the myocardium. Lower two rows show examples where the DL-based denoising approach introduced a false defect (marked by red arrows). The normal-dose images (second column) are also shown for reference. Rows 1, 2, and 4 are images from a male patient, and row 3 is from a female patient.*

Fig. 4 shows representative sample images prior to and after the denoising operation. These images are at 20% low-dose level. On initial inspection, it did appear that the denoising operation yielded images that looked less noisy and more similar to the normal-dose image. This was also reflected in the performance with fidelity-based FoMs. However, on closer inspection, we observed findings that contradicted the initial inspection. In the upper two rows, where actual defects were present (marked by yellow arrows), the DL-based denoising approach removed defects. Note that defects were still observed in the normal and the low-dose images prior to denoising. Similarly, in the lower-two rows, when no defect was present, the DL-based denoising approach created false defects (marked by red arrows). These observations were consistent with the results obtained from the quantitative task-based evaluation studies.

Fig. 5 shows the defect regions, intensity profiles of defect regions, and $\Delta \bar{v}$ obtained from normal dose images and images at low dose before and after applying the DL-based denoising approach with four types of defects at 10% and 5% dose levels. For clearer interpretation, we also looked at the intensity profile along the line at the center of this image. We observed a reduction in $\Delta \bar{v}$ after the DL-based denoising approach was applied, explaining the lower AUC obtained with that approach. The first four columns of Fig. 5 show that the DL-based denoising operation decreased the difference in the means between defect-absent and defect-present cases compared to when no denoising was applied. These observations were also true for 20% and 15% dose levels, as shown in the supplementary material (Fig. S I).

Based on the results of Fig. 5 and the fact that the DL-based denoising approach reduces $\Delta \bar{\bar{f}}$, we expected that it would also reduce the coefficients $\alpha$, and this was indeed observed, as shown in the supplemental results (Fig. S II). This has the impact of degrading observer performance. However, as we observe in Fig. 6, the denoising operation reduced the eigenvalues of the covariance matrices for each defect type. This then has the impact of improving observer performance. In other words, the impact of the denoising operation on the difference in signal means and on the covariance contradict each other. The combined impact of both these effects is shown in Fig. 7. Here, we observe that for each of the defect types, the observer SNR decreased after applying the denoising operator. This shows that while the considered DL-based approach positively impacted the covariance of the noise, overall, it did not improve the defect-detection performance since it reduced the difference in the means between the defect present and defect absent cases.

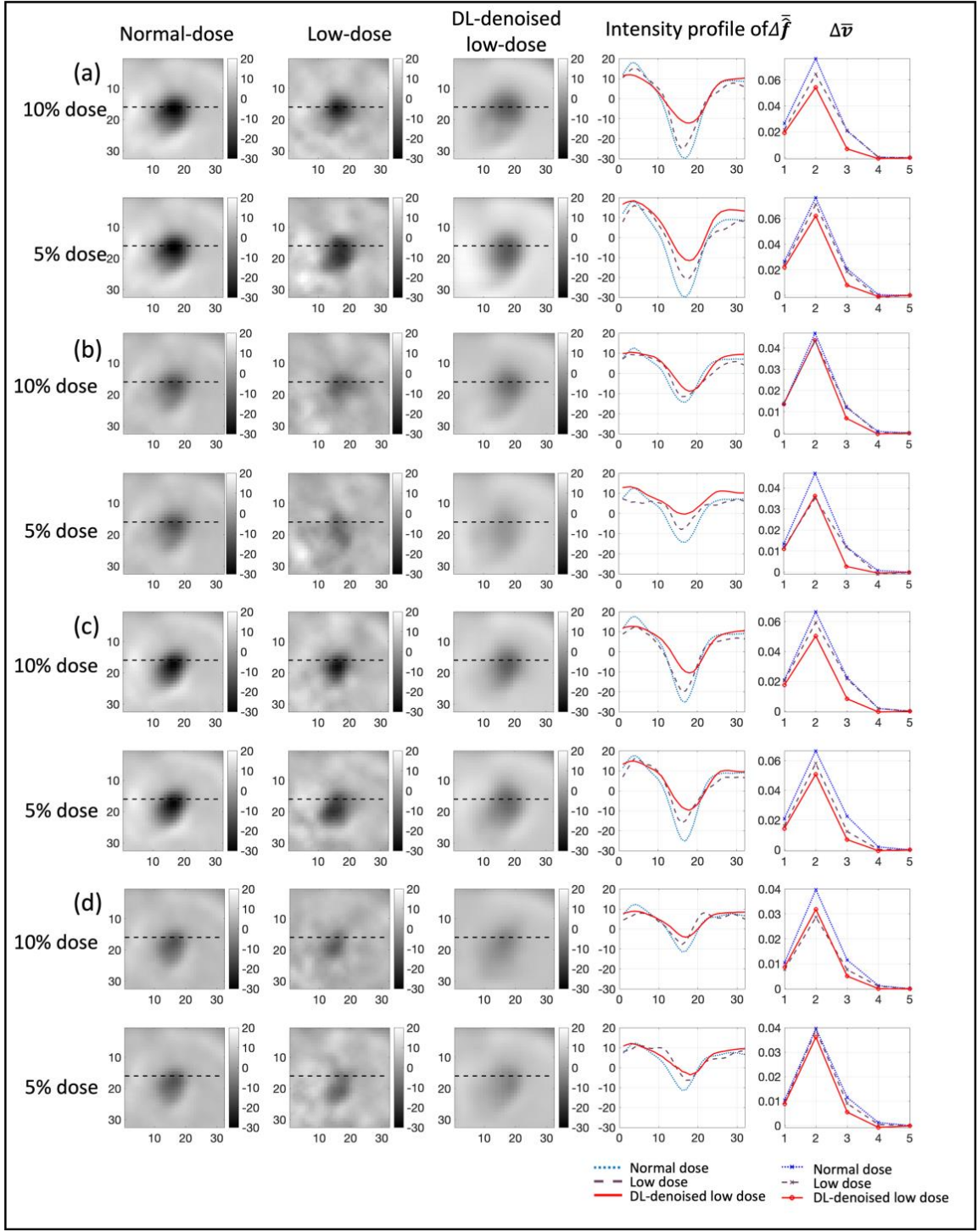

Figure 5: The defect regions, intensity profiles of $\Delta \bar{\bar{f}}$, and $\Delta \bar{v}$ obtained at normal dose and the images at low dose before and after applying the DL-based denoising approach with defect (a) type 1, (b) type 2, (c) type 3, and (d) type 4, at 10% and 5% dose levels. The first three columns show the windowed $\Delta \bar{\bar{f}}$ for the normal dose, the images at low dose before and after applying the DL-based denoising approach. The fourth column shows the intensity profile of $\Delta \bar{\bar{f}}$ corresponding to the dashed line in the defect region. The fifth column shows the profile of $\Delta \bar{v}$ for the corresponding $\Delta \bar{\bar{f}}$.

These observations also align with our visual observations of the impact of DL-based denoising, since the denoising approach does make the image look visually less noisy. However, in this process, when the defect is present, it also reduces the defect contrast. This may be analogously compared to the impact of a low-pass filtering operation on noisy images, which, while making the images look less noisy, may also make the defect less detectable. These observations provide insights into the reasons for the limited performance of the DL-based denoising approach on the defect-detection task.

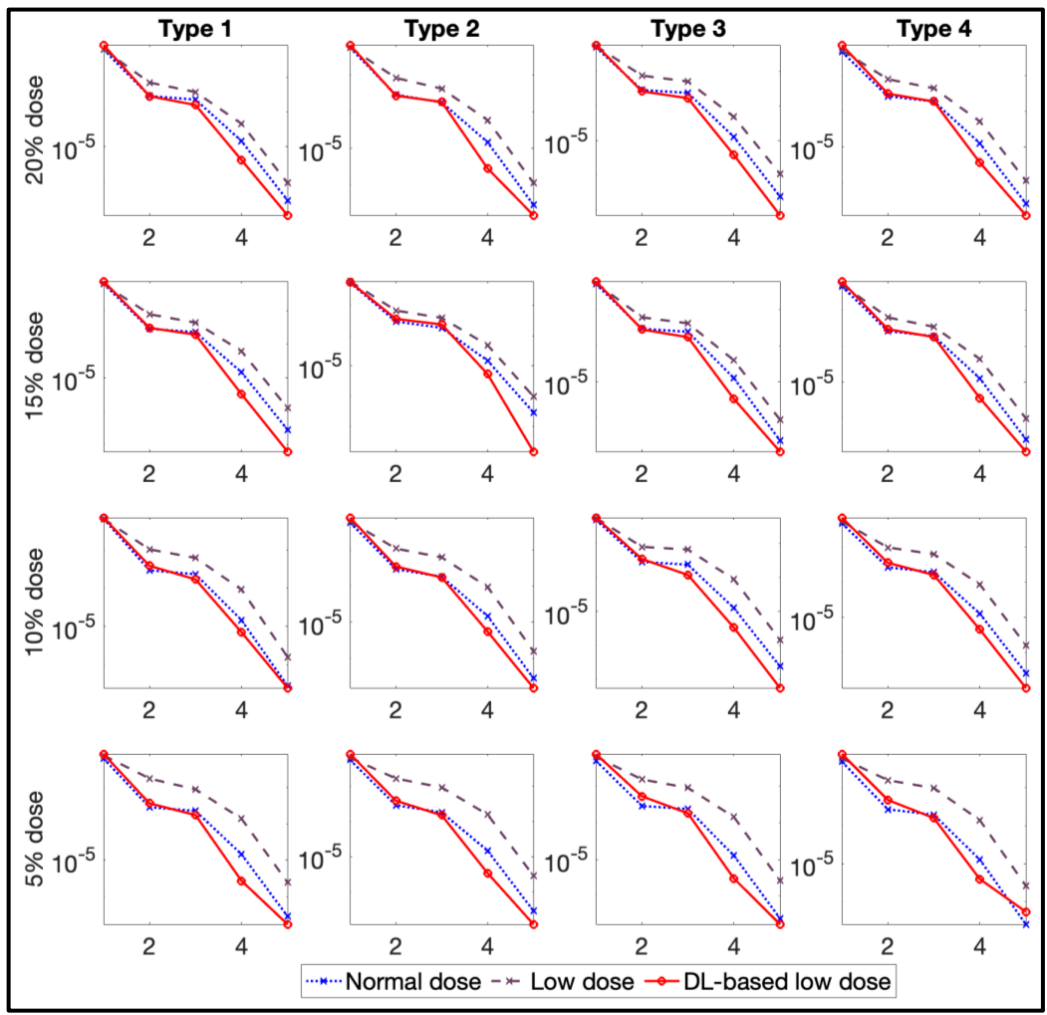

Figure 6. The eigenvalue spectra of the covariance matrix obtained at normal dose and the images at low dose before and after applying the DL-based denoising approach with four types of defects at four dose levels.

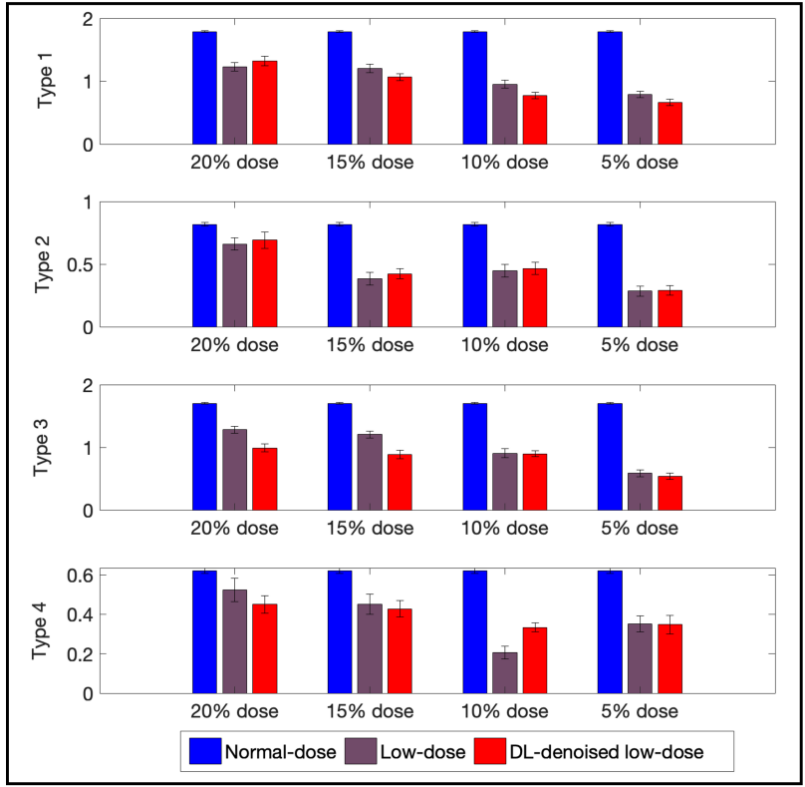

*Figure 7. The observer SNR obtained at normal dose and the images at low dose before and after applying the DL-based denoising approach with four types of defects at four dose levels. Confidence intervals were indicated by black lines.*

# DISCUSSION

In this study, we evaluated a DL-based approach for denoising MPS images acquired at multiple low-dose levels using fidelity-based FoMs and using an observer study on the clinical task of cardiac defect detection. The fidelity-based FoMs suggested that the DL-based denoising approach results in significantly superior performance at all the low-dose levels. Overall, the denoised image looked smoother than the normal-dose images. However, when evaluated on the task of defect detection, the DL-based denoising approach did not yield any improvement in performance. Further, while the images generated by the DL-based denoising approach appeared less noisy, the approach decreased overall contrast in the denoised images, introduced false defects, and removed the true defects in many cases. These results demonstrate the limitation of using fidelity-based FoMs to evaluate DL-based methods and illustrate the importance of evaluating these methods on specific clinical tasks. Such task-based evaluation becomes even more important since DL-based denoising methods that use neural-network-based architectures operate using rules derived from training data. These rules may not be physically interpretable, and thus the correlation with task performance may not be intuitive.[31]

We provided a theoretical analysis to study the impact of post-processing operations on signal-detection tasks for Hoteling-template-based observers. The analysis provides a mechanism to quantify the impact of the post-processing operation on the first and second-order statistics of images. Our observations show that the considered DL-based denoising approach decreased the mean difference between feature vectors of defect-present and defect-absent cases, while also decreasing the eigenvalues of the covariance matrix of the feature vectors. Overall, the combined impacts of these effects decreased the observer SNR. These observations suggest that the key reason for the degraded performance of the considered DL-based denoising approach on the defect-detection task is the reduction in defect contrast, even if the noise properties improve. We emphasize here that the provided theoretical analysis is applicable for investigating the impact of any post-processing operations, including other DL-based methods, on signal-detection tasks with observers that use the first and second-order statistics of the images.

Another objective of this study was to demonstrate the utility, feasibility, applicability, and advantages of task-based evaluation of DL-based approaches *in silico* using VCTs. DL-based approaches are showing tremendous promise in medical imaging, but there is an important need for evaluation of these methods for clinical translation. While such evaluation should ideally be performed clinically, that is very time-consuming, expensive, poses risks to the patients (in terms of exposure to radiation dose), and suffers from multiple logistical challenges such as the need to recruit patients. Thus, there is an important need for strategies to identify only the most promising methods for further clinical evaluation. In this context, VCTs offer an accelerated, scalable, safe, and inexpensive mechanism to identify such promising methods.[28] VCTs do not require patient recruitment. Further, they are scalable since a large population of patients can be generated, which may be required for training the DL network and conducting rigorous observer studies. Next, VCTs do not require subjects to be scanned and thus have minimal imaging costs and risks. Another important feature that VCTs offer is knowledge of ground truth, which further facilitates the rigors of the observer study. Moreover, since the VCTs are conducted with model observers, including anthropomorphic numerical observers, the image interpretation strategy lends itself to highly reproducible findings.

The findings in our study are also consistent with another recent clinical study.[62] In that study, it was observed that a DL-based CT-less attenuation and scatter compensation (ASC) approach for whole-body PET with $^{18}$F-fluorodeoxyglucose yielded images similar to those obtained from a CT-based ASC approach, as per the fidelity-based FoMs of normalized RMSE, PSNR, and SSIM. However, it was observed that the DL-based denoising approach yielded false-negative results due to blurring or missing lesions, and false positives due to pseudo low-uptake patterns.

The findings of this study are not meant to suggest that DL-based methods are broadly ineffective. In fact, in another ongoing project, we are showing the efficacy of such methods on transmission-less ASC in SPECT on clinical defect-detection tasks in MPS.[63] The presented study has several limitations that limit drawing conclusions about the efficacy of DL-based denoising approaches. As mentioned previously, our studies were simulation-based, it is possible that clinical images may contain features that assist the DL-based denoising methods and that are absent in simulated images. This limitation could be addressed by conducting an evaluation study with patient data. The second limitation is that we evaluated the DL-based denoising approach at four dose levels. There may be dose levels for which the investigated DL-based denoising approach works well. Another limitation is that we did not evaluate a wide spectrum of DL-based denoising methods. We chose an approach that was commonly used for the denoising application in SPECT, and, even though our approach was optimized, it is possible that another DL-based denoising[64] approach may be more suitable for this application. For example, our choice of the loss function was the commonly used mean square error between denoised and reference images.[24,58,65,66] However, it is possible that a different loss function, for example, one that accounts for the task from the image,[57,67,68] may yield improved performance. Since our goal was to specifically investigate the need for task-based evaluation of DL-based denoising methods, evaluating a broad array of DL-based denoising methods in a range of study designs and applications using different observers would be beyond the scope of this paper. However, our study motivates the evaluation of these DL-based denoising methods and the development of new DL-based denoising methods in this direction. Given these limitations, we do not draw any conclusions on the efficacy of DL-based denoising approaches.

## CONCLUSION

This study demonstrates that evaluation of a commonly used deep learning (DL)-based denoising method using conventional fidelity-based figures of merit (FoMs) can yield discordant interpretation compared to objective evaluation on clinical tasks in myocardial perfusion SPECT. The results motivate the evaluation of DL-based denoising methods using objective task-based evaluation studies. Our theoretical analysis revealed that the major reason for the degraded performance was that the denoising method reduced the mean difference between defect-absent and defect-present cases. Further, we see that such an evaluation can be conducted computationally and relatively inexpensively within a virtual clinical trial framework and thus help identify promising methods for subsequent human-observer studies. Overall, the study demonstrates the need for and feasibility of task-based evaluation of AI-based methods and provides a theoretical framework to probe the effect of post-processing operations on signal-detection tasks.

## SOFTWARE AVAILABILITY

Software to conduct this study is available at https://github.com/YuZitong/Need-for-Objective-Task-based-Evaluation-of-DL-Based-Denoising-Methods. Pending necessary permissions, we will be disseminating this software widely. Software to assist with objective task-based evaluation studies is also available at multiple other locations, such as the University of Arizona Image quality toolbox, the Metz ROC software at the University of Chicago,[69] the iMRMC Statistical tools hosted at U.S. Food and Drug Administration[70,71] and the Duke Center for Virtual Imaging Trials.[72,73]

## ACKNOWLEDGEMENTS

This work was supported in part by National Institute of Biomedical Imaging and Bioengineering of National Institute of Health (NIH) under grant number R21-EB024647, R01-

EB031051, R01-EB031051-02S1, and R01EB031962. Support is also acknowledged from the NVIDIA GPU grant. We also thank the Washington University Center for High Performance Computing for providing computational resources. The center is partially funded by NIH grants 1S10RR022984-01A1 and 1S10OD018091-01.

The authors thank Paul Segars (paul.segars@duke.edu) and Duke University for providing the XCAT phantom and Kyle Myers, PhD and Mark Anastasio, PhD for helpful discussions.

# CONFLICT OF INTEREST STATEMENT

No potential conflicts of interest relevant to this article exist.